\documentclass[refere]{aa} 
\usepackage{txfonts}
\usepackage[utf8x]{inputenc}
\usepackage{graphicx}
\usepackage{natbib}
\bibpunct{(}{)}{;}{a}{}{,} 
\usepackage{threeparttable}
\usepackage{amssymb}
\title{The nature of the multi-wavelength emission of 3C 111}

\author{S. de Jong\inst{1}, V. Beckmann\inst{1} \and F. Mattana\inst{1}}
\institute{
 Fran\c{c}ois Arago Centre, APC, Universit\'e Paris Diderot, CNRS/IN2P3, CEA/Irfu, Observatoire de Paris, Sorbonne Paris Cit\'e, 10 rue Alice Domon et L\'eonie Duquet, 75205 Paris Cedex 13, France\\
 \email{dejong@in2p3.fr}\label{inst1}}
\date{Accepted 13 July 2012}

\begin{document}

\abstract{}{
We attempt to determine the nature of the high energy emission of the radio galaxy 3C 111, by distinguishing between the effects of the thermal and non-thermal processes.
}
{We study the X-ray spectrum of 3C 111 between 0.4 keV and 200 keV, and its spectral energy distribution, using data from the Suzaku satellite combined with INTEGRAL, Swift/BAT data, and Fermi/LAT data. We then model the overall spectral energy distribution by including radio and infrared data.}
{The combined Suzaku, Swift and INTEGRAL data are represented by an absorbed exponentially cut-off power-law with reflection from neutral material with a photon index $\Gamma = 1.68 \pm 0.03$, a high-energy cut-off $E_\mathrm{cut} = 227_{-67}^{+143} \rm \, keV$, a reflection component with $R=0.7 \pm 0.3$
and a Gaussian component to account for the iron emission-line at 6.4~keV with an equivalent width of $EW = 85 \pm 11 \rm \, eV$. The X-ray spectrum appears dominated by thermal, Seyfert-like processes, but there are also indications of non-thermal processes. The radio to $\gamma$-ray spectral energy distribution can be fit with a single-zone synchrotron-self Compton model, with no need for an additional thermal component.}
{We suggest a hybrid scenario to explain the broad-band emission, including a thermal component (iron line, reflection) that dominates in the X-ray regime and a non-thermal one to explain the spectral energy distribution.}

\maketitle
\section{Introduction}
Radio galaxies are a subclass of active galactic nuclei (AGNs) and in the unification model of AGN radio galaxies are the radio-loud counterparts of Seyfert galaxies \citep{Antonucci1993,Urry1995}. For both of these classes, it is thought that the inclination is large, such that the observer views the core, which is dominated by thermal processes through absorbing material.
This differents from blazars, where the inclination angle is very small. The observed emission originates in the relativistic jet, which is dominated by non-thermal processes. Owing to the Doppler boosting of the emission in the jet, the energy of the observed emission from blazars can reach the $\gamma$-ray regime, even up to TeV energies.  
Several non-blazar AGNs 
 have also been found to emit significantly in the $\gamma$-ray regime \citep{Hartman2008, Abdo2010,Ackermann2011}. Although there have been several theories (i.e. misaligned jet, shocks in the radio lobes) about the origin of this high-energy radiation, conclusive support of any of these theories is yet to be found.\\
3C~111 is a nearby \citep[z=0.049,][]{Sargent1977} flat-spectrum FR-II radio galaxy \citep{Fanaroff1974, Linfield1984} that displays strong, broad emission-lines in the optical \citep{Sargent1977} and an iron-emission line in the X-ray regime \citep{Lewis2005}, similar to Seyfert galaxies. In the radio regime, the galaxy's emission is dominated by the relativistic jet, which has an angle of 18\textdegree \, to our line of sight \citep{Jorstad2005}. The projected size of the jet is 78 kpc \citep{Bridle1984}. There is no visible counter-jet, although a bright lobe is detectable in the opposite direction of the observed jet, which is likely fed by the undetected counter-jet \citep{Linfield1984}. \\
Earlier studies of 3C~111 in the X-ray band have identified the high energy cut-off of the spectrum.
\citet{Dadina2007} found a lower limit of $E_\mathrm{cut} \geq 82 \rm \, keV$ using {\it BeppoSAX} data and \citet{Ballo2011} derived a similar lower limit to the cut-off of $E_\mathrm{cut} \geq 75 \rm \, keV$, using data from XMM-\textit{Newton} and {\it Suzaku}/XIS and PIN. \citet{Molina2009}, using \textit{XMM-Newton}, {\it Swift}, and {\it INTEGRAL} data, constrained the cut-off energy to be $E_\mathrm{cut}=126 ^{+193}_{-50} \rm \, keV$.
It is uncertain whether a reflection component is present in the X-ray spectrum. An upper limit of $R\leq 2.25$ was found by \citet{Dadina2007} using {\it BeppoSAX} data and constraints of $R=0.9 \pm 0.6$ were derived by \citet{Molina2009} and $R=0.35\pm0.06$ by \citet{Ballo2011} using {\it Suzaku}/XIS and PIN data. \citet{Ballo2011} used {\it XMM-Newton} data to find the reflection in the energy range 0.4--10 keV to be $R=0.19^{+0.05}_{-0.04}$. 
\citet{Rivers2011} did not detect a reflection component, using {\it RXTE} data between 3 keV and $\gtrsim$ 100 keV. 
The equivalent width (EW) of the iron line at 6.1 keV has been measured several times and is likely variable. \citet{Ballo2011} and \citet{Tombesi2010} found similar values of $EW = 75 \pm 13 \rm \, eV$ ({\it Suzaku}/XIS and PIN) and $EW = 86 \pm 16 \rm \, eV$ ({\it Suzaku}/XIS), respectively. Using {\it Suzaku}/XIS data of three different observations taken two years later, \citet{Tombesi2011} found a smaller $EW$ of between $EW > 33 \rm \, eV$ and $EW > 40 \rm \, eV$. 
Similarly, \citet{Ballo2011} used {\it XMM-Newton} data to fit the 0.4--10 keV spectrum and found an $EW = 38^{+11}_{-9} \rm \, eV$. \\
3C~111 has been detected in $\gamma$-rays by {\it CGRO}/EGRET \citep{Hartman1999, Sguera2005,Hartman2008}, and was included in the first {\it Fermi}/LAT catalogue \citep{Abdo2010firstsourcecatalog}. In the second {\it Fermi}/LAT catalogue, the source was omitted since it had no longer been significantly detected \citep{Fermisecondsourcecat}. 3C~111 is likely variable in the $\gamma$-ray regime \citep{Ackermann2011}. Using 24 months of {\it Fermi}/LAT data, \citet{Grandi2012} found 3C~111 to be detectable during a short time-period ($\Delta t \sim 30-60$ days), limiting the radius of the emission region to be $R<0.1$ pc (assuming a Doppler factor of $\delta =3$) based on to causality arguments. The detectability of the source in the GeV energy range coincided with an increase in the flux in the millimeter, optical and X-rays regimes, indicating the emission is likely to emerge from the same region. Since the outburst in millimeter, optical, and X-rays is associated with the ejection of a bright radio knot, this indicates that the GeV emission originates from the radio core within 0.3 pc of the central supermassive black hole.
\\
To understand the physical processes causing the emission in 3C~111, we first study the X-ray to $\gamma$-ray spectrum to find whether the emission is the product of thermal (Seyfert-like) or non-thermal processes (blazar-like) or a combination of both. We then model the spectral energy distribution (SED) to understand the processes dominating the broad-band radiation of 3C~111 from radio to high energies.
\section{Observations and data reduction}
To cover a large energy band from X-rays to $\gamma$-rays and study the high-energy emission from 3C~111, we used data from several different instruments.
\subsection{Suzaku-XIS/PIN}
3C~111 was observed by Suzaku from 22 August 2008 to 25 August 2008 with a total elapsed time of 236.9 ks in HXD nominal pointing mode. We analysed data from the X-ray Imaging Spectrometer \citep[XIS,][]{Koyama2007} and Hard X-ray Detector \citep[HXD,][]{Takahashi2007}. The XIS instrument consists of three separate CCD detectors and has an energy range of $\sim0.2-12.0$ keV. The HXD is a collimated instrument, consisting of two independent detector systems, silicon PIN diodes that function in the range $\sim10-60$ keV and GSO scintillation counters covering the energy range $\sim30-600$ keV.\\
For the observations made with the XIS, we used the clean events provided by the instrument team where the standard event cuts had been applied.\\
The flux of 3C~111 is too low above 70 keV to extract a significant spectrum from the HXD/GSO detector, hence we used only HXD/PIN data. 
The Suzaku team provides a response file for the PIN, which depends on both the epoch in which the data were taken and the pointing mode. In addition a non X-ray background (NXB) per observation is available for the PIN analysis. 
After filtering the source and background files in time using the good time intervals, we extracted the source and background spectra. The source spectrum needed to be corrected for dead-time events, which we achieved using \textsc{hxddtcor}. The NXB was simulated with ten times more counts to suppress statistical errors. We therefore needed to increase the exposure time of the background spectrum by a factor of ten. Using \textsc{xspec}, we simulated the cosmic X-ray background (CXB) using the PIN response file for the flat emission distribution in the proper epoch and point mode. We combined the NXB and CXB files  
and used this as a background in our analysis.
\subsection{INTEGRAL IBIS/ISGRI}
The International Gamma-Ray Astrophysics Laboratory (INTEGRAL; \citet{Winkler2003}) is a $\gamma$-ray observatory with several instruments on board.
We used the INTEGRAL Soft Gamma-Ray Imager (ISGRI), which is part of the Imager on Board INTEGRAL Spacecraft (IBIS), a coded-mask detector. ISGRI is sensitive between 15 keV to 1 MeV \citep{Lebrun2003}. \\
We used all data collected by INTEGRAL since the launch of the satellite up to August 2009, with a total exposure time of 508 ks. We first created individual spectra for all science windows (a science window is all data produced during one pointing) using the \textsc{ibis\_science\_analysis} routine. We then summed all the individual spectra to achieve a higher signal-to-noise ratio.
\subsection{Swift-BAT}
The Burst Alert Telescope (BAT) aboard the Swift satellite is a coded-aperture camera with an energy range of 14--195 keV for imaging \citep[][]{Barthelmy2005}. Since the BAT telescope monitors the sky continuously and has a large field of view (1.4 sr, partially coded), it has regularly observed 3C~111. \\
We used data from the 58-month hard X-ray survey, from 2004 December to the end of 2009 May \citep[see][]{Tueller2010,Baumgartner2010}. The detected flux for 3C~111 in this survey is $1.2 \times 10^{-10}  \, \rm erg \, cm^{-2} \, s^{-1}$ between 14 keV and 195 keV.
\subsection{Fermi-LAT}
The Large Area Telescope (LAT) aboard the Fermi satellite operates in an energy range between 20 MeV and 300 GeV \citep{Atwood2009}. The LAT is a pair-conversion telescope with a very wide field of view that scans the sky continuously. During the nominal all-sky survey observing mode of {\it Fermi}/LAT, a total effective exposure time of 83.7 Ms was accumulated in the direction of 3C~111 between 4 August 2008 and 20 April 2011. \\
We selected diffuse-event-class photons (P6\_V3 instrument response functions) between 100 MeV and 200 GeV in a circular region of interest with radius of 15\textdegree \, around the source. Events with a zenith angle of more than 105\textdegree \, were excluded \citep{Abdo2009}. We used the standard cuts proposed by the Fermi team based on the data quality of the events and the instrument configuration.\\
The maximum-likelihood analysis tool \textsc{gtlike} models all the emission in a given region, which can contain several sources. The goodness of fit is expressed as the log-likelihood value which is the probability of obtaining the data given an input model. After creating a model including all detectable sources in the field, we used \textsc{gtlike} to fit the model until the log-likelihood value was maximised. We found the test statistic value of 3C~111 to be $TS=12.8$, which corresponds to a significance of $\sim 3 \sigma$. The data provided only two significant points so it is impossible to fit the spectrum in e.g. \textsc{xspec}. However, when creating the spectrum, \textsc{gtlike} gives a spectral slope of $\Gamma_\gamma=2.41\pm 0.17$ across the range 0.1--200 GeV and a flux of $f=1.2 \times 10^{-8} \, \rm ph \, cm^{-2} \, s^{-1}$. 
\section{Results}
\subsection{X-ray spectrum}

\begin{figure}[!h]
\includegraphics[width=9cm, keepaspectratio=true]{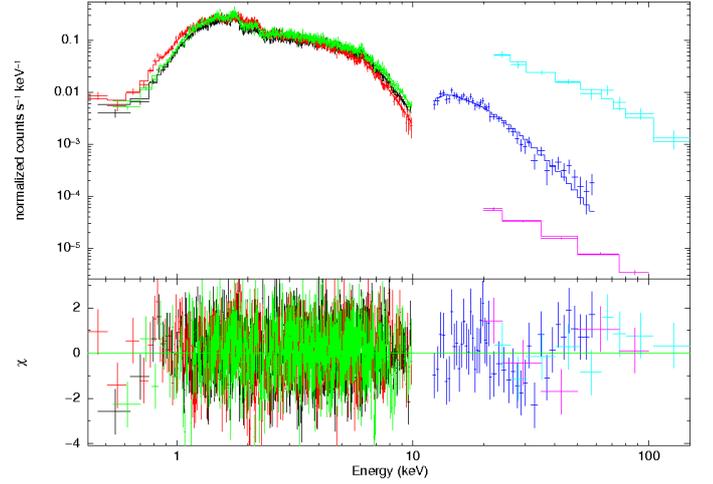}
\caption{The count spectrum of the combined Suzaku/XIS (0.4-10 keV), Suzaku/PIN (12-60 keV), INTEGRAL ISGRI (20-200 keV), and Swift/BAT (15-150) data with the fitted model: an absorbed cut-off power-law with reflection from neutral material and a Gaussian component to account for the iron line at 6.4 keV. The bottom panel shows the residuals in terms of the standard deviation with error bars of size one sigma.}
\label{suzaku_swift_integral_fit}
\end{figure}

\begin{figure}[!h]
\includegraphics[width=7cm, keepaspectratio=true]{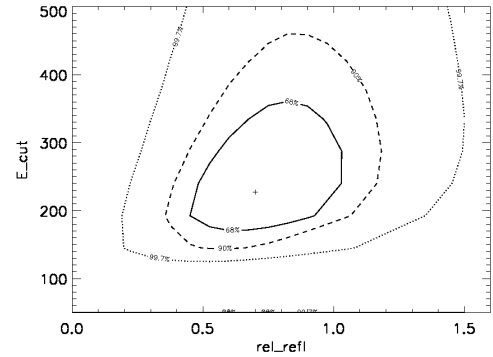}
 \caption{Error contours for the best-fit {\tt pexrav} model, reflection component $R$ vs the high-energy cut-off $E_\mathrm{cut}$. The contour levels correspond to $\Delta \chi^2$ of the 68\%, 90\%, and 99.7\% statistical confidence regions. A cut-off below 130 keV is excluded at a 99.7\% level.}
\label{errorcontour}
\end{figure}

We simultaneously fitted the X-ray spectra between 0.4 keV and 200 keV, in \textsc{xspec} \citep{Arnaud1996}. 
We started our analysis by fitting the spectrum with an absorbed power-law and a Gaussian component at 6.1~keV to account for the redshifted iron emission-line. The fit resulted in a reduced chi-squared of $\chi_\nu^2 =1.14$ (1678 d.o.f.). We tried to improve our fit by using a cut-off power-law, which yielded a better fit with $\chi_\nu^2=1.12$ (1677 d.o.f.). 
We explored the possibility of a broken power-law, but with a $\chi_\nu^2=1.14$ (1676 d.o.f.), this gave no improvement in the fit. Adding a reflection component did further improve the fit, $\chi_\nu^2 =1.10$ (1676 d.o.f., F-test probability $4 \times 10^{-8}$). The resulting fit and the residuals are shown in Figure \ref{suzaku_swift_integral_fit}. We derived the following best-fit parameters and 90\% errors: the value for the equivalent hydrogen column-density $N_{\rm H} = (9.0 \pm 0.2) \times 10^{21}$ cm$^{-2}$, a power-law index of $\Gamma =  1.68 \pm 0.03$, a high-energy cut-off  $E_\mathrm{cut} = 227_{-67}^{+143}  \rm \, keV$ and a reflection scaling factor of $R = 0.7 \pm 0.3$. 
Figure \ref{errorcontour} shows the confidence contours of the reflection component versus the cut-off energy with $\Delta \chi^2$ of 68\%, 90\%, and 99.7\%. We can exclude a cut-off below 130 keV at the 99.7\% confidence level.\\ 
We fitted the data with a more physical model, {\tt compPS}  \citep{Poutanen1996}, which describes the process of thermal Comptonization. In this model, seed photons from the cool, thick accretion disc 
are injected into the electron plasma. The electron cloud can have several geometries, and we chose a plane-parallel slab: we also applied other geometries but found that they neither influenced the parameter values nor improved the fit. The electrons in the cloud have a Maxwellian distribution with an electron temperature $T_\mathrm{e}$ and an optical depth $\tau$ related to the Compton parameter $y = 4 \tau kT_\mathrm{e} / (m_\mathrm{e} \, c^2)$. The seed photons are up-scattered from their initial energy $E_i$ to $E_f = e^y E_i$. The Compton scattering of the seed photons with the electrons in the plasma results in a spectrum that is afterwards reflected from a cool medium and then smeared out by the rotation of the disc.\\
The best-fit model gave a $\chi_\nu^2 =1.10$ (1676 d.o.f.). The values that we found for this model are an equivalent hydrogen column-density $N_{\rm H} = (9.0 \pm 0.2 ) \times 10^{21}$ cm$^{-2}$, a temperature of the electrons $k T_\mathrm{e} = 91 _{-48}^{+22} \rm \, keV$, a reflection component $R = 1.8 ^{+ 0.5}_{-0.7}$, and a Compton parameter of $y = 0.6 \pm 0.1$, which corresponds to an optical depth $\tau = 0.8$. \\ 
We also fitted all data sets individually with an absorbed power-law and a Gaussian component. The values of the flux levels and power-law indices for these individual best fits to the data can be found in Table \ref{fitvaluespl}.  \\
The measurements of the iron line at 6.4 keV are consistent in all three {\it Suzaku}/XIS spectra. After adding the three spectra together, we derived a line energy $E_\mathrm{line} = 6.11 \pm 0.02 \rm \, keV$ and an $EW = 85 \pm 11 \rm \, eV$.

\begin{table*}
\centering
\caption{The parameters for the individual power-law fits of the data, and their 90\% confidence levels.
 }
{\footnotesize
\vspace*{0.5cm}
\begin{tabular*}{\textwidth}{l|l|l|c|l|l}  
\hline \hline
\noalign{\smallskip}

Instrument & Epoch & Exposure time & PL index & $f \, \rm [erg \, cm^{-2} \, s^{-1}]$ & Energy range \\ 
\hline
\noalign{\smallskip}
XIS0  & 22-25 August 2008  & 95.4 ks & 1.60 $\pm$ 0.02 & $(2.20 \pm 0.02) \times 10^{-11}$ & 0.4--10 $\rm keV$ \\ 
XIS1  & 22-25 August 2008  & 95.4 ks & 1.59 $\pm$ 0.02 & $(2.24^{+0.03}_{-0.02})\times 10^{-11}$ & 0.4--10 $\rm keV$ \\ 
XIS3  & 22-25 August 2008  & 95.4 ks & 1.63 $\pm$ 0.02 & $(2.2 \pm 0.02) \times 10^{-11}$ & 0.4--10 $\rm keV$ \\ 
PIN   & 22-25 August 2008  & 101.9 ks & 1.52 $\pm$ 0.14 & $(4.8 \pm 0.4) \times 10^{-11}$ & 12--60 $\rm keV$ \\ 
ISGRI & 24 March 2003-19 August 2009 & 508 ks & 1.90 $\pm$ 0.20 & $(1.2 \pm 0.2 )\times 10^{-10}$ & 20--200 $\rm keV$ \\ 
BAT   & December 2004-May 2009 & 54 months$^a$& 1.99 $\pm$ 0.09 & $(1.05\pm0.05) \times 10^{-10}$ & 15--150 $\rm keV$ \\
LAT   & 4 August 2008-20 April 2011 & 83.7 Ms & 2.4 $\pm$ 0.2 & $(6\pm2)\times 10^{-12}$ & $>$ 100 $\rm MeV$ \\

\noalign{\smallskip}
\hline \hline
\end{tabular*}
\label{fitvaluespl}
\tablefoot{
\tablefoottext{a}{Elapsed time} 
}
}
\end{table*}

\subsection{Fitting of the time-averaged SED}
\begin{figure}[!h]

\includegraphics[width=8cm, keepaspectratio=true]{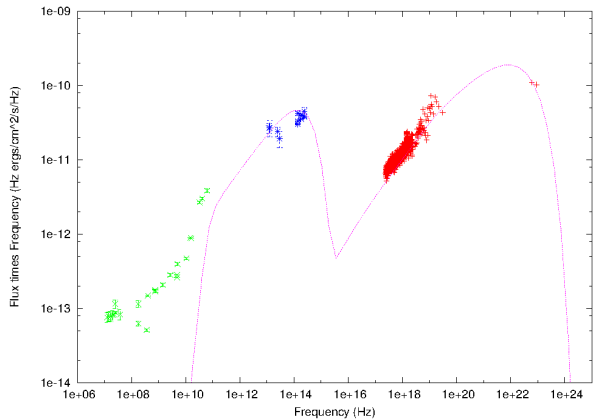}
 \caption{SED showing 3C~111 unabsorbed fluxes and the SSC one-zone model. Crosses indicate data extracted and analysed in this work. We have added archival and deabsorbed IR and radio points from NED. The line shows the SSC model. See Table \ref{sedparameters} for the values used.}
\label{sed_fit}
\end{figure}

After we had created a SED based on our data, we modelled it with a public synchrotron self-Compton (SSC) code developed by \citet{Krawczynski2004}\footnote{http://jelley.wustl.edu/multiwave/spectrum/?code}. The SSC mechanism assumes an isotropic population of high-energy electrons that emit synchrotron radiation followed by inverse Compton scattering of the synchrotron photons to higher energies \citep{Maraschi1992}. The inverse Compton component may also include a thermal component of external seed photons (e.g. from the broad-line region or accretion disc) that are also Compton upscattered to high energies \citep[external Compton component,][]{Dermer1993}. The electron population is located in a spherical volume of radius $R_{\mathrm s}$ with a randomly orientated magnetic field $B$ that moves relativistically towards the observer with a bulk Lorentz factor of $\Gamma$ and an angle between the jet and the line of sight of $\theta$. Thus, the radiation is Doppler-shifted with a Doppler factor $\delta=[\Gamma(1-\beta \cos \theta)]^{-1}$. The electron energy spectrum in the jet-frame follows a broken power-law, with indices $p_1$ and $p_2$, and is characterised by a minimum and maximum energy ($E_\mathrm{min}$, $E_\mathrm{max}$) and a break energy ($E_\mathrm{br}$). The power in the SED is mostly influenced by the Doppler factor and both the radius and magnetic field of the emitting region. The shapes of the synchrotron and inverse Compton peaks and the peak frequencies depend strongly on the energies that define the electron population. \\ 
Since we only had data in the X-ray and $\gamma$-ray regimes, it was impossible to model the entire SED. We therefore added archival radio data, neglecting the observations that only included the core, and infrared data from the NED archive\footnote{http://ned.ipac.caltech.edu/}. Assuming that the infrared emission is produced in the same region as the X-rays (the accretion disc or inner jet), we were able to use the column density found in the X-ray spectra, $N_{\rm H} = 9.0 \times 10^{21} \rm \, cm^{-2}$ to calculate the extinction in the V-band $A_{\mathrm V} = N_{\rm H}/(1.79 \times 10^{21} \rm \, cm^{-2})$ \citep{Predehl1995}. From the extinction in the V-band, we calculated the extinction in the near-infrared bands J, H, and K using the correction factors from \citet{Schlegel1998}. We removed the {\it Suzaku}/XIS data below 1 keV to avoid possible contamination by starburst emission.\\
Using our best fit of the X-ray spectrum, we were able to constrain some of the parameters. The indices of the power-law are tied to the power-law index of the X-ray spectrum: the first power-law index is $p_1=2$ (below $E_\mathrm{br}$). 
 Starting with this parameter and the initial parameters of the code, we optimised the model. We focused on the energies that characterise the electron population to define the shape of the peaks, and to change the overall energy output in the SED we adjusted the radius of the emitting region and its magnetic field, and the Doppler factor.\\
We found that the break in the electron power-law is insignificant and we therefore used a single power-law with index $p_1= p_2$. Furthermore, we found the minimum energy of the electron distribution to be $E_\mathrm{min}=5.6 \times 10^{6}\rm \, eV$ and the maximum energy $E_\mathrm{max}=6.8 \times 10^{9} \rm \, eV$.
 The Doppler factor is $\delta= 14$, the radius of the spherical emission-volume is $R_{\mathrm s} =2 \times 10^{16} \rm \, cm$, and the magnetic field is $B=4 \times 10^{-4} \rm \, G$. The parameters that we used to fit the data can be found in Table \ref{sedparameters}, and the resulting plot is presented in Figure \ref{sed_fit}. The addition of an external Compton component does not improve the model fit.

\begin{table*}

\centering

\caption{Parameters used to model the SED of 3C~111 using the code by \citet{Krawczynski2004}. For comparison, we include the fit parameters for the core of Centaurus A \citep{AbdoCenAcore2010} and Markarian 421 \citep[low flux state,][]{Blazejowski2005}.
 }
{\footnotesize
\vspace*{0.5cm}
\begin{tabular*}{\textwidth}{@{\extracolsep{\fill}}l|l|l|l|l} 
\hline \hline
\noalign{\smallskip}

Symbol  & Description & 3C 111 & Cen A & Mrk 421\\ 
\hline
\noalign{\smallskip}
$\delta$ & Doppler factor & 14  & 1.0 & 10.0 \\
$R_{\mathrm s}$ & Radius of the emission volume $[\rm cm]$ & $2 \times 10^{16}$ &  $3.0\times10^{15}$ & $7.0\times10^{15}$\\
$B$ & Magnetic field of emission volume $[\rm G]$ & $4\times 10^{-2}$ &  6.2 & 0.405 \\
$E_\mathrm{min}$ & Minimum energy of the electron distribution  $[\rm eV]$& $ 5.6 \times 10^{6}$ &  $1.5\times10^8$& $3.2 \times 10^6$ \\
$E_\mathrm{max}$ & Maximum energy of the electron distribution  $[\rm eV]$& $6.8 \times 10^{9}$  & $5.0 \times 10^{13}$ & $1.7\times10^{11}$\\
$E_\mathrm{br}$ &  Break energy of the electron distribution  $[\rm eV]$ & -  &  $4.0 \times 10^8$& $2.2 \times10^{10}$\\
$p_1$ &  Spectral index of electron spectrum ($E_\mathrm{min}$ to $E_\mathrm{br}$) & 2 & 1.8 & 2.05 \\
$p_2$ &  Spectral index of electron spectrum ($E_\mathrm{br}$ to $E_\mathrm{max}$) & - &4.3 & 3.6\\
$\log L$ & Bolometric luminosity (3-200 keV)  $[\rm erg \, s^{-1}]$ & 44.7 & 42.5$^a$ & 45.5$^b$ \\ 

\noalign{\smallskip}
\hline \hline
\end{tabular*}
\label{sedparameters}\\
\tablefoot{
\tablefoottext{a}{
 \citet{Beckmann2011};}
\tablefoottext{b}{ \citet{Lichti2008}} \\
}
}
\end{table*}

\section{Discussion}
To evaluate the physical processes that dominate the high-energy emission from X-ray to $\gamma$-rays in 3C~111, we studied the X-ray spectrum and the broad-band SED.
\subsection{X-ray spectrum}
\begin{figure}[!h]
 \includegraphics[width=8cm, keepaspectratio=true]{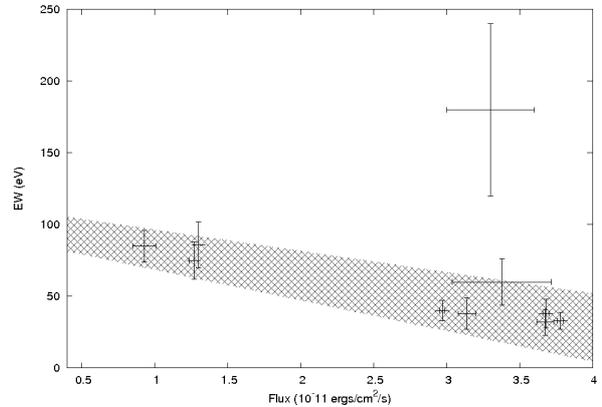}
\caption{Flux of the source versus the EW of the iron line. Data taken from \citet{Tombesi2010,Tombesi2011},
 \citet{Ballo2011}, \citet{Rivers2011}, \citet{Lewis2005}, and \citet{Eracleous2000} and this work are plotted as points with error bars. We used a linear regression fit to derive a correlation between the two parameters. The fit is plotted with grey dashed lines indicating the $1\sigma$ error region.}
\label{ewflux}
\end{figure}

We constrained the high-energy cut-off of 3C~111 by applying the {\tt pexrav} model, which describes an exponentially cut-off power-law spectrum reflected from neutral material \citep{pexrav1995} in {\sc xspec}. We found a cut-off $E_\mathrm{cut} = 227_{-67}^{+143}  \rm \, keV$ using data from {\it INTEGRAL}/ISGRI, {\it Suzaku}/XIS and PIN, and {\it Swift}/BAT. An indication of the cut-off can already be seen in the power-law indices of the individual fits, which go from $\Gamma \sim1.6$ in the soft X-rays to $\sim1.9$ in the hard band (Table \ref{fitvaluespl}). We also applied the physical {\tt compPS} model to the same data, which yielded an electron temperature $k T_\mathrm{e} = 91 _{-48}^{+22} \rm \, keV$ that can be related to the cut-off energy via $E_{\mathrm cut} \simeq 3kT_{\mathrm e}$. The electron temperature that we found using the {\tt compPS} model is consistent with the cut-off energy that we derived using the {\tt pexrav} model. \\
The value we found for the high-energy cut-off is consistent with the upper limits reported by \citet{Dadina2007} and \citet{Ballo2011} and the measurement by \citet{Molina2009}. 
\citet{Rivers2011} found, using data from {\it RXTE} from 3 keV to $\gtrsim 100 \rm \, keV$, that the addition of a cut-off does not improve their fit of 3C~111. This might be because we used a combination of different instruments and a wider energy range for the spectrum presented here. \\
A high-energy cut-off is a typical property of the high-energy spectra of Seyfert galaxies, and the value we derived for 3C~111 is in the expected range for cut-off energies observed in Seyferts \citep{Beckmann2009,Dadina2008}. \\
We measured the reflection in 3C~111 using both the {\tt compPS} and the {\tt pexrav} model to be $R = 1.8 ^{+ 0.5}_{-0.7} $ and $R = 0.7 \pm 0.3$, respectively. Since these models are very different ({\tt pexrav} is a descriptive model and {\tt compPS} is a physical model), we expect to find different measures of the reflection. \citet{Beckmann2011} applied both models to the hard X-ray spectrum of the radio galaxy Cen~A and found that the reflection in the case of {\tt compPS} is also slightly higher though consistent with the reflection found using the {\tt pexrav} model.\\
The reflection scaling factor we derived, $R=0.7 \pm 0.3$, using the {\tt pexrav} model can be compared to earlier work using the same model. Our results are consistent with an upper limit reported by \citet{Dadina2007}, as well as with the constraints obtained by \citet{Molina2009} and \citet{Ballo2011}. The latter also reported a significantly lower result, $R=0.19^{+0.05}_{-0.04}$, when using only soft X-ray data between 0.4 keV and 10 keV. This is to be expected since the reflection component depends on the energy band used to observe it. In addition \citet{Rivers2011} did not detect a reflection component when applying the {\tt pexrav} model between 3 keV and $\gtrsim$~100 keV. They constrained the model so as to avoid a high-energy cut-off, which might explain the difference from our result.\\
The value we measured is also consistent with the average Seyfert properties values of $R=1.2^{+0.6}_{-0.3}$ for Seyfert 1 and $R=1.1^{+0.7}_{-0.4}$ for Seyfert 2 galaxies \citep{Beckmann2009}. \\
For the iron line at 6.11 keV, we found that its $EW = 85 \pm 11 \rm \, eV$, which is similar to the results of \citet{Ballo2011} and \citet{Tombesi2010}. Smaller values between $EW > 33 \rm \, eV$ and $EW > 40 \rm \, eV$ were reported by \citet{Tombesi2011}. Since the source is variable, it is likely that the iron line EW varies depending on the continuum. Figure \ref{ewflux} shows the correlation between the flux between 4 keV and 10 keV and the EW of the iron line for several observations. The correlation between the flux and the EW has a significance of $> 95\%$, a Pearson's test coefficient of $r = -0.3$, and a Spearman's rank correlation coefficient $r_s = -0.7$, showing a believable correlation between the flux level and the EW. When we did not consider the large EW measurement based on {\it RXTE} data presented by \citet{Rivers2011}, we get a highly significant correlation with a probability of $>99\%$. 
\citet{Tombesi2010,Tombesi2011} also report the power-law indices changing for the different data sets. Similarly, using a different data set, \citet{Ballo2011} derived an $EW = 38^{+11}_{-9} \rm \, eV$. Since there is a difference of several months between both data sets this value can again be explained by the source variability. 

We found that the EW is smaller than the average value for Seyfert galaxies \citet{Dadina2008}, at $EW = 448 \pm 67 \rm \, eV$, which can be compared to the average for Seyfert 1 galaxies of $EW = 222 \pm 33 \rm \, eV$ and for Seyfert 2 galaxies of $EW = 693 \pm 195 \rm \, eV$. 
The correlation between EW of the iron line and the underlying power-law flux indicates that the line flux does not vary significantly, whereas the continuum varies by a factor of $\sim 5$. This can be naturally explained if the majority of the continuum flux is produced in a non-thermal and variable process. The small value of the EW of the iron line may indicate that the X-ray spectrum of 3C~111 does not only contain a thermal, but also a non-thermal contribution.
\\
We also fitted the spectrum with the physical {\tt compPS} model, deriving a temperature for the electron cloud $k T_\mathrm{e} = 91 _{-48}^{+22} \rm \, keV$, a Compton parameter $y = 0.6 \pm 0.1$ (which corresponds to an optical depth of $\tau = 0.8$), and a reflection component $R = 1.8 ^{+ 0.5}_{-0.7} $. \\
NGC~4151 is a well-studied case of a Seyfert core at hard X-rays. \citet{Beckmann2005} fitted the {\tt compPS} model to the spectrum of the Seyfert galaxy NGC~4151 across the range 2-300 keV. They found for the electron plasma temperature $kT_\mathrm{e} = 94^{+4}_{-10} \rm \, keV$, a reflection component $R = 0.72 \pm 0.14$ and an optical depth of $\tau=1.3^{+0.13}_{-0.05}$. The hard X-ray spectrum (3--1000 keV) of Centaurus~A, which is another nearby $\gamma$-ray detected radio galaxy, was also modelled with {\tt compPS} by \citet{Beckmann2011}. This yielded an electron plasma temperature of $kT_\mathrm{e} = 206 \pm 62 \rm \, keV$, a Compton parameter $y = 0.42^{+0.09}_{-0.06}$ (which corresponds to an optical depth of $\tau = 0.26$), and a reflection component of $R = 0.12^{+0.09}_{-0.10}$. 
NGC~4151 is optically thick, whereas both Cen~A and 3C~111 are not. The optical thickness of NGC~4151 depends on the flux state and can decrease to $\tau = \sim 0.3-0.6$ in the dim state \citep{Lubi2010}, when the electron temperature can increase to $T_\mathrm{e} \sim 180-230 \rm \, keV$. The optical depth of 3C~111 is similar to that of the dim state of NGC~4151, but the electron temperature is closer to the electron temperature of the brighter state.  The reflection measured in 3C~111 is higher than those found in both NGC~4151 and Cen~A.\\
\citet{Chatterjee2011} concluded that the X-ray spectrum of 3C~111 is of thermal inverse Compton origin, based on a correlation between the optical and X-ray fluxes, as well as a weak optical polarisation and a smaller variance in the optical than X-ray flux on shorter timescales. They concluded that these findings are consistent with a reprocessing model where the X-rays are mostly produced by inverse Compton scattering of thermal optical/UV seed photons from the accretion disc. \\
The cut-off we measured using the {\tt pexrav} model may have either a thermal or a non-thermal origin. It can be interpreted as a high-energy cut-off measured in Seyfert galaxies. In the case of 3C~111, the cut-off value is also comparable to the typical cut-offs found in Seyfert galaxies. The cut-off observed in the {\tt pexrav} model can also be caused by non-thermal inverse-Compton scattering processes. In this case, the smooth turn-over of the inverse Compton branch towards higher energies appears as a cut-off in the {\tt pexrav} model. While in the thermal inverse-Compton case the cut-off is exponential, the non-thermal spectrum is curved in such a way that $\gamma$-ray and VHE emission cannot be excluded. It is impossible to differentiate between an exponential and a simple cut-off in the hard X-rays based on the data at hand, thus both the non-thermal and the thermal interpretations of the curved X-ray spectrum are still valid.
The reflection component is a property of Seyfert galaxies and a result of thermal processes. The value of the one we detect is in the typical range for reflection in Seyfert galaxies. 
The EW of the observed iron line is smaller than expected for Seyfert galaxies and is also variable, implying that the continuum is similarly variable. We therefore suggest that, while the thermal processes in the X-ray band in 3C~111 seem to dominate, we cannot exclude a non-thermal contribution.
\subsection{$\gamma$-rays}
3C~111 was proposed as a counterpart of the $\gamma$-ray source 3EG~J0416+3650 in the third {\it CGRO}/EGRET catalogue, even though it fell outside the 99\% probability region \citep{Hartman1999}. A re-analysis of the data \citep{Sguera2005} concluded that 3EG J0416+3650 is likely associated with 3C~111. There are no other plausible counterparts in the EGRET error region, which is also larger than previously thought 
because the quoted errors were statistical only and did not take into account the larger systematic errors caused by inaccuracies in the Galactic diffuse model.
The EGRET data were re-analysed \citep{Hartman2008} when it was found that 3EG~J0416+3650 is likely the result of the blending of more than one source. One of these components (detected only above 1~GeV) can be associated with 3C~111. 
Furthermore, 3C~111 was included in the first Fermi/LAT source catalogue \citep{Abdo2010firstsourcecatalog} with a significance of $4.3 \sigma$. In the second {\it Fermi}/LAT source catalogue, 3C~111 was excluded because the source was no longer significantly detected \citep{Fermisecondsourcecat}. However, 3C~111 is very likely to be variable \citep{Ackermann2011} and therefore no longer detectable in the second year. This was also confirmed by  \citet{Kataoka2011}, who found a significance $>$5$\sigma$ for 3C~111 using 24 months of {\it Fermi}/LAT data and by \citet{Grandi2012} who found that the $\gamma$-ray emission is not persistent, but flaring and associated with the ejection of bright radio knots.\\
Since 3C~111 was detected by {\it CGRO}/EGRET and {\it Fermi}/LAT at different epochs, we assumed that the source is a variable $\gamma$-ray emitter. We included data from the first {\it Fermi}/LAT catalogue where the source was detected significantly. Our analysis gives a power-law index similar to that given in the first catalogue with a comparable flux level \citep{Abdo2010firstsourcecatalog}.
\subsection{Spectral energy distribution}
To construct the SED of a variable source, simultaneously acquired data are necessary in order to ensure that we make an unbiased measurement. The data used in the SED of 3C~111 were not acquired simultaneously, which can affect the results in the sense that the SED is composed of measurements from different spectral states.
This effect is more severe for very variable objects, whereas at most wavebands 3C~111 is moderately variable. We refer, for example to \citet{Beckmann2007} who found no significant variability in hard X-rays for their study of 9 months of  {\it Swift}/BAT data. For the radio domain, \citet{Grossberger2012} present decade-long light curves showing variations of up to a factor of 2 which we can consider as insignificant variability in the context of the SED.
For other radio galaxies, it has also been shown that the use of non-simultaneous data to construct the SED does not introduce any significant bias. For example, the radio galaxies NGC~1275 \citep{Anton2004} and Pictor~A \citep{Brown2011} were also similarly analysed using time-averaged SEDs. The results for the physical parameters should thus not be significantly affected by the moderate variability of 3C~111.
\\
The archival radio data that we used to model the SED were not well-represented by our SSC model. For observatories that operate at different wavebands, there are large differences in the data acquired in terms of the field of view and resolution. The resolution depends on the wavelength via $RL = \lambda /D$, where $RL$ is the resolution, $\lambda$ the wavelength, and $D$ the dish diameter. It is therefore possible that we probe different regimes in the broad-band SED, something that the one-zone model does not account for. \\
In SSC models, the relativistic electron-energy distribution is often assumed to be a broken power-law with a break energy $E_\mathrm{br}$, a power-law index before the break of $p_1 < 3$ and a power-law index after the break of $p_2 > 3$ \citep[see for example][]{Ghisellini1996, Tavecchio1998, Krawczynski2004}. The electron population is confined to region of radius $R_{\mathrm s}$ and a magnetic field $B$ that moves with a Doppler factor $\delta$ along the jet. The electrons emit synchrotron radiation and inverse Compton radiation using the synchrotron photons as seed photons. 
For electrons with energies below the break energy $E_\mathrm{br}$, cooling of the electrons is the dominant emission source and in the regime after the break, the escape of electrons from the source dominates. The peak synchrotron power is emitted by electrons at the break energy \citep{Tavecchio1998}. In the case of 3C~111, we found that using a broken power-law gave a break energy of $E_\mathrm{br}=1.6 \rm \, GeV$ and power-law indices of $p_1 = 2$ and $p_2 = 2.2$. The break between the two power-laws is insignificant. We therefore applied a single power-law with an index $p = 2$, based on the power-law index of the X-ray spectrum. In the case of a single power-law, the maximum synchrotron power is emitted near the maximum energy of $E_\mathrm{max} \simeq 7  \rm \, GeV$.\\
The energies that define the electron distribution, $E_\mathrm{min}$ and $E_\mathrm{max}$ influence the shape of the synchrotron and inverse Compton peak and were chosen by ourselves empirically. 
If the minimum energy, $E_\mathrm{min}$, decreases, the maximum of the peak occurs at lower frequency and the slopes are less steep. Increasing the $E_\mathrm{min}$ increases the depth between the synchrotron and inverse Compton peak and steepens the rise and fall of the peaks. The maximum energy, $E_\mathrm{max}$, is tied to the peak frequencies of both the synchrotron and inverse Compton peak: increasing this value would increase the peak frequencies. 
\\
\citet{AbdoCenAcore2010} modelled the SED of the core of the radio galaxy Cen~A, using simultaneous data from the radio up to the $\gamma$-ray regime, applying a single-zone SSC. 
Comparing the parameters for this model and our own (see Table~\ref{sedparameters}), it is clear that the models have significantly different parameter values. 
 The magnetic field has a much higher value in the case of Cen~A ($B = 6.2 \rm \, G$), than the value we found for 3C~111 ($B =0.04 \rm \, G$). The higher magnetic field increases the resulting flux because the synchrotron power depends on the magnetic field. In contrast, the Doppler factor used to model Cen~A is low ($\delta =1$) compared to the value we found for 3C~111, $\delta = 14$. A lower Doppler factor means that the emission is less boosted and therefore appears less energetic. 
 There are also three orders of magnitude difference between the emission volumes: for Cen~A, $R_{\mathrm s} = 3 \times 10^{15} \rm \, cm$ was used, whereas we assumed $R_{\mathrm s} =  2 \times 10^{16}  \rm \, cm$ for 3C~111. Since this defines the amount of particles this also causes the flux to be lower for Cen~A than for 3C~111. The higher value used for the magnetic field is attenuated by the lower Doppler factor and radius for Cen~A, which causes the overall output to be lower, as expected for the less luminous source.\\
Pictor~A is another FR-II radio galaxy that was detected by {\it Fermi}/LAT as reported by \citet{Brown2011}, who modelled the SED of one of the hot-spots in the radio-lobe of Pic~A with a one-zone SSC model using the same code that we applied. They concluded that it is impossible to describe both the X-ray and $\gamma$-ray emission using the SSC model. Since the $\gamma$-ray emission is very variable it is likely to originate from the jet, whereas the X-rays originate from the hot-spot. This appears to indicate that more than one zone is needed to model the entire SED.\\
\citet{Blazejowski2005} also used the Krawczynski code to model the high-energy peaked BL~Lac object Mrk~421, assuming only an SSC component. The parameter values they derived are shown in Table~\ref{sedparameters}. Both the Doppler factor ($\delta =10$) and the emitting region radius ($R_{\mathrm s} = 7 \times 10^{15} \rm \, cm$) chosen for Mrk~421 are smaller than those chosen for 3C~111, but are larger than those of Cen~A. Similar to Cen~A, the magnetic field value for Mrk~421, $B =0.4 \rm \, G$, is a factor of 10 stronger than 
 for 3C~111. The smaller radius and Doppler factor decrease the flux, but owing to the strong magnetic field the flux in the SED for Mrk~421 is higher than the flux of 3C~111.\\
We conclude that the overall emission from 3C~111 can be modelled with a simple synchrotron self-Compton model, where no additional thermal Compton component is needed. Since the X-rays appear to be (mostly) of thermal origin, the SED provides an upper limit to the non-thermal emission in the X-ray band. There may be a thermal component in the SED, but with the current data we are unable to disentangle the possible thermal component from the overall non-thermal emission.
\section{Conclusion}
The origin of the high-energy emission from non-blazar AGNs remains unclear. \citet{Marscher2002} suggested that with the acceleration of the inner regions of the accretion disc a shock front will stream along the jet and the expulsion of a superluminal bright knot will follow. \citet{Grandi2012} show, using {\it Fermi}/LAT data, that the GeV emission of 3C~111 appears to originate from a compact knot confined to within 0.1 pc. This knot is clearly separate from the core and placed 0.3 pc from the central engine.\\
By analysing the X-ray spectrum and the broad-band SED, we have studied the nature of the high-energy emission of the radio galaxy 3C~111.
We have presented an X-ray spectrum between 0.4 keV and 200 keV using data of 3C~111 acquired by several instruments and showed that the best-fit model is an absorbed cut-off power-law with both a reflection component and a Gaussian component to account for the iron line at 6.4 keV. The values we found for the reflection and high-energy cut-off are similar to those found in Seyfert galaxies, which would indicate that there is a thermal core visible. The cut-off can also originate from non-thermal processes and the EW of the iron line is variable and smaller than expected for Seyfert galaxies. We therefore conclude that the X-ray spectrum is mainly of thermal origin, but there may be a small non-thermal contribution. \\
Using the X-ray spectrum, together with $\gamma$-ray data from {\it Fermi}/LAT and archival deabsorbed radio and infrared data, we modelled the broad-band SED of 3C~111 using a single-zone synchrotron self-Compton model. This model is non-thermal and we also did not need to include an additional thermal component to model the SED. Since the X-ray emission is likely to have a combined thermal and non-thermal origin, the SSC model we used may overestimate the non-thermal contribution in the X-ray band and should therefore be considered an upper limit. \\
In conclusion, it seems that the high-energy emission from 3C~111 consists of both thermal and non-thermal components. In the X-ray spectrum, the thermal components manifest themselves in terms of an iron line and reflection. The non-thermal component is visible through the variability in the EW of the iron line. The high-energy cut-off can be the result of either thermal or non-thermal inverse Compton scattering, but our present spectrum does not allow us to distinguish which process is occurring. The broadband SED can be modelled with a non-thermal model, but it is possible there is a thermal component that we are unable to discern with the current data set. 
\begin{acknowledgements} 
Based on data provided by INTEGRAL, an ESA project funded by ESA member
states (especially the PI countries: Denmark, France, Germany, Italy, Spain,
Switzerland), Czech Republic, Poland, and with the participation of Russia and
the USA. We acknowledge the use of public data from the Swift data archive. This research has made use of the NASA/IPAC Extragalactic Database (NED), which is operated by the Jet Propulsion Laboratory, California Institute of Technology, under contract with the National Aeronautics and Space Administration. This research has made use of NASA's Astrophysics Data System Bibliographic Services. We thank the anonymous referee for the comments that helped to improve the paper. 
\end{acknowledgements}
\bibliographystyle{aa} 
\bibliography{3C111.bib} 

\end{document}